\documentclass[nofootinbib,longbibliography,prd,floatfix,superscriptaddress,twocolumn]{revtex4-1}
\usepackage[english]{babel}
\usepackage{amsmath,amssymb,amsbsy,amstext,amsthm,booktabs,simplewick,exscale,relsize,graphicx,amsfonts,upgreek,slashed,color,multirow,dcolumn,bm,enumerate,url,hyperref}
\newcommand{\toolfont}[1]{\texttt{#1}}

\begin{document}

\title{Dark Kinetic Heating of Neutron Stars and An Infrared \\ Window On WIMPs, SIMPs, and Pure Higgsinos}

\author{Masha Baryakhtar}
\affiliation{Perimeter Institute for Theoretical Physics, Waterloo, Ontario, N2L 2Y5, Canada}

\author{Joseph Bramante}
\affiliation{Perimeter Institute for Theoretical Physics, Waterloo, Ontario, N2L 2Y5, Canada}

\author{Shirley Weishi Li}
\affiliation{CCAPP and Department of Physics, The Ohio State University, Columbus, OH, 43210, USA}

\author{Tim Linden}
\affiliation{CCAPP and Department of Physics, The Ohio State University, Columbus, OH, 43210, USA}

\author{Nirmal Raj}
\affiliation{Department of Physics, University of Notre Dame, Notre Dame, IN, 46556, USA}

\begin{abstract} We identify a largely model-independent signature of dark matter interactions with nucleons and electrons. Dark matter in the local galactic halo, gravitationally accelerated to over half the speed of light, scatters against and deposits kinetic energy into neutron stars, heating them to infrared blackbody temperatures. The resulting radiation could potentially be detected by the James Webb Space Telescope, the Thirty Meter Telescope, or the European Extremely Large Telescope. This mechanism also produces optical emission from neutron stars in the galactic bulge, and X-ray emission near the galactic center, because dark matter is denser in these regions. For \mbox{GeV - PeV} mass dark matter, dark kinetic heating would initially unmask any spin-independent or spin-dependent dark matter-nucleon cross-sections exceeding $2 \times 10^{-45}$ cm$^2$, with improved sensitivity after more telescope exposure. For lighter-than-GeV dark matter, cross-section sensitivity scales inversely with dark matter mass because of Pauli blocking; for heavier-than-PeV dark matter, it scales linearly with mass as a result of needing multiple scatters for capture. Future observations of dark sector-warmed neutron stars could determine whether dark matter annihilates in or only kinetically heats neutron stars. Because inelastic inter-state transitions of up to a few GeV would occur in relativistic scattering against nucleons, elusive inelastic dark matter like pure Higgsinos can also be discovered.
\end{abstract}

\maketitle

Despite ongoing searches, the identity of dark matter remains a mystery. Terrestrial detectors looking for dark matter impinging on known particles have found no dark sector scattering events in up to a hundred kilogram-years of data. While some dark matter models have been excluded by these searches, many well-motivated candidates remain untested. Earthbound direct detection is considerably less sensitive to dark matter that couples to Standard Model particles primarily through inelastic or spin-dependent interactions, as well as dark matter much heavier or lighter than the nuclear mass of silicon, argon, germanium, xenon, or tungsten.

This letter demonstrates that the aggregate impact of dark matter falling onto neutron stars results in thermal emission detectable with imminent telescope technology. Detecting or constraining dark matter using nearby neutron stars requires dedicated search strategies and observation times a few orders of magnitude beyond standard surveys. In addition, locating an old neutron star within fifty parsecs of Earth, where $\sim 100$ old neutron stars reside, may be critical to near-future searches for dark kinetic heating. We will show that such efforts are warranted by the extraordinary sensitivity dark kinetic heating has for a broad variety of dark matter models. 

A compelling insight developed in the remainder of this document is that any appreciable dark matter interactions with Standard Model particles will heat neutron stars, through the deposition of kinetic energy that dark matter gains falling into steep neutron star gravitational potentials. This dark kinetic heating of neutron stars depends only on the total mass of accumulated dark matter, and is therefore sensitive to dark matter masses spanning dozens of orders of magnitude. 
As a consequence, dark kinetic heating of neutron stars provides a powerful complement to, and indeed could surpass, terrestrial direct detection searches for dark matter interactions. This approach can be compared to prior work on dark matter in compact stars, which examined dark matter that heats white dwarfs and neutron stars by annihilating in their cores, $e.g.$ \cite{Kouvaris:2007ay,Bertone:2007ae,Kouvaris:2010vv,deLavallaz:2010wp,McCullough:2010ai,Bramante:2017xlb}. 

{\em{\bf 1.} Dark kinetic heating.} The flux of dark matter through a neutron star depends upon the maximum impact parameter for which dark matter in the halo intersects a neutron star with mass $M$ and radius $R$ \cite{Goldman:1989nd},
$
b_{\rm max} = \left(\frac{2 G M R}{v_{\rm x}^2} \right)^{1/2}  \left(1-\frac{2 G M}{R} \right)^{-1/2},
$
where $G$ is Newton's constant, and $v_{\rm x}$ is the velocity of the dark matter.
The total mass rate of dark matter passing through the neutron star is then
\begin{align}
\dot{m} = \pi b_{\rm max}^2 v_{\rm x} \rho_{\rm x},
\end{align}
where $\rho_{\rm x}$ is the ambient density of dark matter. Using a best-fit dark matter density and halo velocity \mbox{$\rho_{\rm x} \sim 0.42 ~{\rm GeV~cm^{-3}}$}, $v_{\rm x} \sim 230~{\rm km~s^{-1}}$ \cite{Pato:2015dua}, we find that for neutron stars near Earth, $\dot{m} \sim 4 \times 10^{25}~{\rm GeV~s^{-1}}$.

The total kinetic energy that can be deposited by dark matter is, to good approximation, given by dark matter's kinetic energy at the surface of the neutron star
\begin{align}
E_{\rm s} \simeq m_{\rm x} \left(\gamma -1 \right),
\end{align}
where for a typical neutron star $\gamma \sim 1.35$. Then the rate of dark kinetic energy deposition is given by
\begin{align}
\dot{E}_{\rm k} = \frac{E_{\rm s} \dot{m}}{m_{\rm x}} f \simeq 1.4 \times 10^{25}~{\rm GeV ~s^{-1}}~\left( \frac{f}{1} \right),
\end{align}
where $f  \in \left[0,1 \right]$ is the fraction of dark particles passing through the star that become trapped in the neutron star interior. This fraction depends on the cross-section for dark matter to scatter against nucleons and electrons. We define a capture probability
\begin{align}
f \equiv {\rm Min} \left[\frac{\sigma_{\rm nx}}{\sigma_{\rm sat}},1\right],
\label{eq:cap}
\end{align}
where $\sigma_{\rm sat} $ is the ``saturation" cross-section, for which all transiting dark matter is captured. Dark matter becomes captured in the neutron star's gravitational potential, if the energy it deposits from scattering with the neutron star exceeds its initial halo kinetic energy far away from the neutron star. We now determine the size and scaling of $\sigma_{\rm sat} $, which will depend on the dark matter mass $m_{\rm x}$. Here we consider dark matter scattering off neutrons -- this treatment can be extended to scattering off electrons and protons in the neutron star \cite{InPrep}.

\begin{figure}[t]
\includegraphics[width=0.48\textwidth]{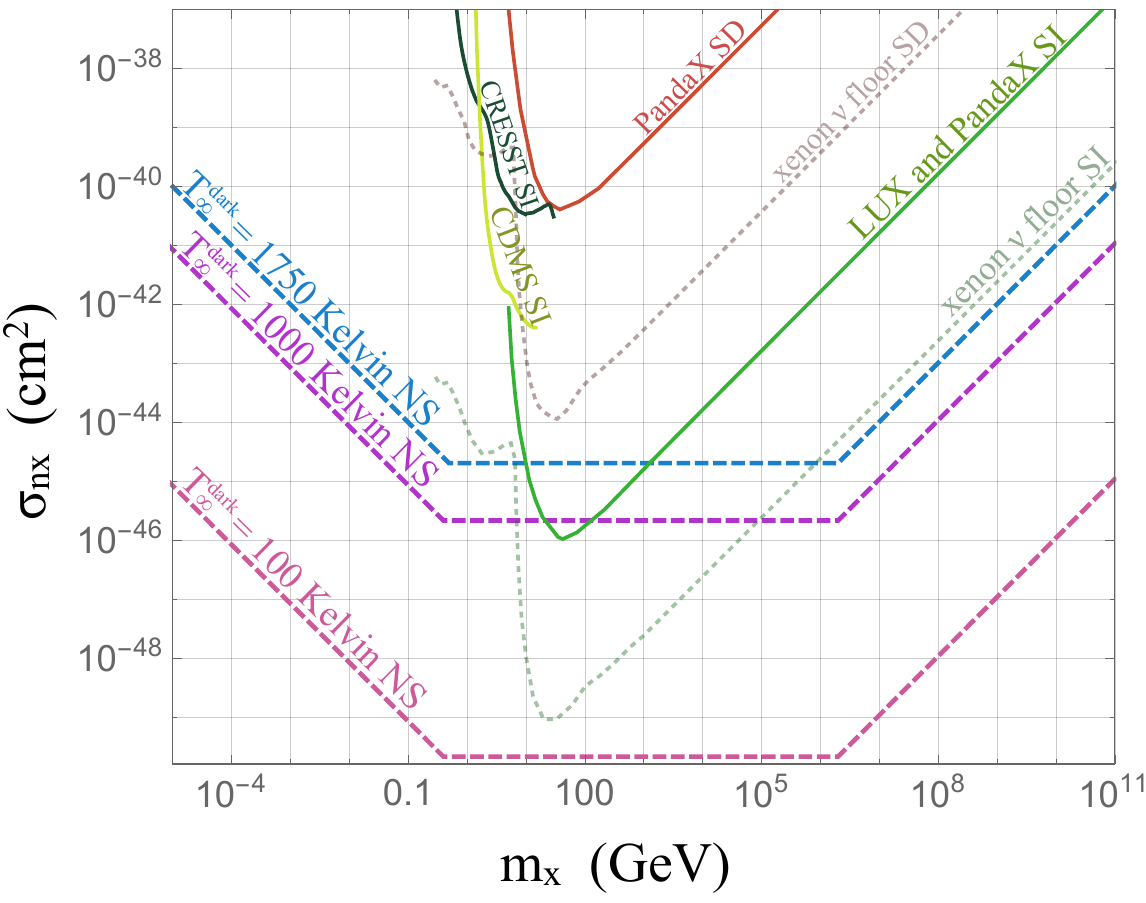}
\caption{Dark kinetic heating sensitivity to dark matter-neutron cross-sections ($\sigma_{\rm nx}$), obtainable from neutron stars near Earth with blackbody temperatures of $100-1750$ Kelvin, are indicated with dashed lines. A $ 1750$ K blackbody temperature is the maximum imparted by dark kinetic heating, for a $1.5~M_{\odot}$, $R=10$ km neutron star that captures the entire flux of dark matter passing through it, for dark matter density $\rho_{\rm x} = 0.42 ~{\rm GeV~cm^{-3}}$ \cite{Pato:2015dua}. While radiation from a $1750$~K neutron star at 10 parsecs could be detected by JWST, TMT, or E-ELT, imaging a $\lesssim 1000$~K neutron star requires future telescopes. As detailed in Sec. 3, old neutron stars cool to $\sim 100$ K after a billion years. Dark kinetic sensitivity curves apply to both spin-dependent (SD) and spin-independent (SI) interactions, since scattering occurs off individual neutrons. Bounds from LUX \cite{Akerib:2016vxi}, PandaX \cite{Tan:2016zwf,Fu:2016ega}, CDMS \cite{Agnese:2014aze}, and CRESST \cite{Angloher:2014myn} are shown, which use $\rho_{\rm x} = 0.3 ~{\rm GeV~cm^{-3}}$, alongside the SD and SI xenon direct detection neutrino floors \cite{Ruppin:2014bra}.}
\label{fig:kindm}
\end{figure}

\emph{{\bf 2a.} Capture for $ {\rm GeV} \lesssim m_{\rm x} \lesssim {\rm PeV}$.} In this mass range, dark matter is captured after scattering once with the neutron star. In the rest frame of the neutron star, a dark matter-nucleon scattering event depletes the dark matter kinetic energy by
\begin{align}
\Delta  E_s = \frac{m_n m_x^2 \gamma^2 v_{\rm esc}^2}{m_n^2 + m_x^2 +2\gamma m_x m_n} \left(1- {\rm cos}~\theta_c \right),
\label{eq:DeltaEs}
\end{align} 
where $v_{\rm esc}^2 \simeq 2G M /R$ is the incoming speed of dark matter, and $\theta_c$ is the scattering angle in the center-of-mass frame. For dark matter masses smaller than $m_{\rm x} \simeq {\rm PeV}$, dark matter becomes bound after one scattering event. This follows from comparing dark matter's initial kinetic energy in the halo, blueshifted in the rest frame of the neutron star, $\gamma m_{\rm x} v_{\rm x}^2/2$, to the kinetic energy lost in an average scatter, $\Delta E_s$. Equating these two quantities, we find that the maximum mass for dark matter captured by one scatter is $m_{\rm x} \sim ~{\rm PeV}$.  Therefore for GeV -- PeV mass dark matter, the minimum cross-section for dark matter to deposit all of its kinetic energy into the neutron star is the per-neutron cross-section for which dark matter scatters once, \mbox{$\sigma_{\rm sat}^{\rm single} \simeq \pi R^2 m_{\rm n}/M  \simeq 2 \times 10^{-45}~{\rm cm^2}~\left(\frac{1.5~{\rm M_{\odot}}}{M} \right)\left(\frac{R}{10~{\rm km}} \right)^2.$}

\emph{{\bf 2b.} Capture for $m_{\rm x} \lesssim {\rm GeV}$.} For dark matter lighter than about a GeV, the saturation cross section increases inversely with dark matter mass as a result of ``Pauli blocking." Because the neutron star is composed of highly degenerate neutrons, protons, and electrons, the probability for dark matter to scatter with these fermions is diminished by the Pauli exclusion principle, which forbids degenerate fermions from becoming excited to a momentum state already occupied by another fermion. This reduces the number of nucleons available for the dark matter to scatter against by a factor $\sim \frac{\delta p}{p_{\rm F,n}}$, where the momentum transferred by the dark matter is $\delta p \sim  \gamma m_{\rm x} v_{\rm esc}$, and a typical neutron Fermi momentum in the neutron star is $p_{\rm F,n} \simeq 0.45~{\rm GeV}~(\rho_{NS} / (4 \times 10^{38} ~{\rm GeV~cm^{-3}}))$ \cite{Shapiro:1983du}. Therefore, the saturation cross-section for sub-GeV mass dark matter is $\sigma_{\rm sat}^{\rm Pauli} \simeq  \pi R^2 m_{\rm n} p_{\rm f}  /( M \gamma m_{\rm x} v_{\rm esc} )$ \mbox{$ \simeq 2 \times 10^{-45}~{\rm cm^2}~\left(\frac{{\rm GeV}}{m_{\rm x}} \right)\left(\frac{1.5~{\rm M_{\odot}}}{M} \right) \left(\frac{R}{10~{\rm km}} \right)^2$.}

\emph{{\bf 2c.} Capture for $m_{\rm x} \gtrsim {\rm PeV}$.} A single scatter will be insufficient to capture dark matter heavier than a PeV. In this case, multiple scatters during dark matter's passage through the star are required for the dark matter to become gravitationally bound. The energy lost after $N \sim n \sigma_{\rm nx} R$ scatters inside the neutron star is approximately $ \Delta E_{\rm s} N \sim\Delta E_{\rm s}  n \sigma_{\rm nx} R$, where $n$ is the number density of neutrons in the neutron star and $N$ is the typical number of scatters for dark matter transiting the neutron star. Because the energy lost in multiple scatters scales linearly with $\sigma_{\rm nx}$, and dark matter's initial halo kinetic energy also scales linearly with $m_{\rm x}$, it follows that at large dark matter masses $m_{\rm x} \gtrsim {\rm PeV}$, the cross-section required for dark matter capture increases linearly with the dark matter mass (see discussion in \cite{Bramante:2017xlb}) so that $\sigma_{\rm sat}^{\rm multi} \simeq   2 \times 10^{-45}~{\rm cm^2}~\left(\frac{m_{\rm x}}{{\rm PeV}} \right)\left(\frac{1.5~{\rm M_{\odot}}}{M} \right) \left(\frac{R}{10~{\rm km}} \right)^2$.

In all above cases, as is evident in Eq.~\eqref{eq:cap}, the amount of dark matter captured and the resulting dark kinetic heating will decrease linearly with cross-sections smaller than $\sigma_{\rm sat}$, because $\dot{E}_{\rm k} \propto \sigma_{\rm nx}$. Minor refinements can be made to these capture calculations \cite{McDermott:2011jp,Bramante:2013hn,Bell:2013xk,Bramante:2013nma}, $e.g.$ accounting for the slightly increased escape velocity in the neutron star interior \cite{Gould:1987ir,deLavallaz:2010wp,Kouvaris:2010jy}, enhanced multiscatter capture for $m_{\rm x}\sim{\rm PeV}$ dark matter particles using a Boltzmann velocity distribution \cite{Bramante:2017xlb}, and capture enhancement from scattering against heavy nuclei in the crust of the neutron star. These depend on the structure of the neutron star crust and degenerate core \cite{Bertoni:2013bsa}, typically yielding percent-level increases to the capture rate \cite{Gould:1991va,deLavallaz:2010wp,Kouvaris:2010vv,Bramante:2017xlb}. 

Upon capture by the neutron star, dark matter rapidly injects its kinetic energy. For dark matter lighter than a GeV, most of its kinetic energy is deposited after a single scatter, as can be verified with Eq.~\eqref{eq:DeltaEs}. Heavier dark matter follows an orbital path that re-intersects the neutron star interior, re-scattering until most of its kinetic energy is deposited \cite{Kouvaris:2010jy}. The energy lost in each transit will be $\Delta E_{\rm t} \sim 2 G m_{\rm n} (R^{-1} - r_{\rm o}^{-1}) (\sigma_{\rm nx} / 10^{-45} {\rm cm^2})$, where $r_{\rm o}$ is the size of the dark matter orbit. The time for dark matter to deposit most of its kinetic energy is short, $t_{\rm dep} \lesssim {10~\rm days} \left( \frac{ 10^{-49} {\rm cm^2}}{\sigma_{\rm nx} } \right) \left(\frac{m_{\rm x} }{ {\rm PeV}}\right) $, with this expression normalized to the longest deposition time for parameters of interest.

\emph{{\bf 3.} Neutron star temperature.} Neutron stars primarily cool through neutrino and photon emission. 
While young neutron star cooling rates can have a strong dependence on the equation of state, neutron stars older than a few million years, $i.e.$ the majority of neutron stars in the Milky Way, have cooling curves of a simple form,
$
dT/dt = (\epsilon_{\rm ISM} +\epsilon_{\rm x}-\epsilon_{\rm \gamma})/c_{\rm v},
$
where $c_{\rm v} = \sum c_{\rm i,v}$ is the total heat capacity with $c_{\rm n,v} \simeq \frac{2 \pi^2 k_B^2 m_n n_n}{3 p_{\rm Fn}^2} T$ and $c_{\rm p,v} \simeq \frac{2 \pi^2 k_B^2 m_{\rm p} n_{\rm p}}{2 p_{\rm Fp}^2} T$ the heat capacity for neutron star fluid with neutron (proton) number density $n_{\rm n}$ ($n_{\rm p}$) \cite{Page:2004fy}, and $\epsilon_{\rm \gamma} = L_\gamma/(4 \pi R^3/3) = 4 \pi \sigma R^2 T^4/(4 \pi R^3/3) $ is the photon emissivity of the neutron star. Additional sources of neutron star thermal energy \cite{Yakovlev:2004iq,Pons:2008fd,Zeldovich:1969aa} include heating from accretion of interstellar matter and dark matter, parameterized by $\epsilon_{\rm ISM}$ and $\epsilon_{\rm x} $, respectively. The blackbody emission outlined here matches the cooling model of \cite{Page:2004fy}, where neutron stars become isothermal and cool to $T \lesssim 10^3$ K within 20 Myr and $T \sim 100$ K after a Gyr \cite{Yakovlev:2004iq}, although this depends on whether there are additional mechanisms that heat old neutron stars. We note that $T \sim 1000$ K is a factor of a hundred below current temperature bounds on neutron stars \cite{Prinz:2015jkd}. One additional putative contributor to neutron star temperature is magneto-thermal heating. However, simulations indicate that magneto-thermal effects damp out after a Myr, independent of initial magnetic field strength \cite{Pons:2008fd}.

Another possible background to dark kinetic heating is interstellar medium accretion onto neutron stars, which depends on their historic paths through the Milky Way disk and halo \cite{1993ApJ...403..690B,2010A&A...510A..23S}. Present-day ISM heating can potentially be discerned from dark kinetic heating, because accreted ions preferentially warm the poles of the neutron star and emit some X-ray photons \cite{1995ApJ...439..849Z,1995ApJ...454..370B}. In practice, neutron star magnetic fields may deflect all but a fraction of incident ISM \cite{2003ApJ...594..936P}. Fortuitously, the local 100 parsecs of ISM (the ``Local Bubble" \cite{2009ApJ...700.1299J}) is relatively underdense, with ISM densities as low as $\sim 10^{-3}~ {\rm GeV~{\rm cm^{-3}}}$, making the present-day dark matter heating contribution predominant in this region.

\begin{figure}[t!]
\includegraphics[width=0.48\textwidth]{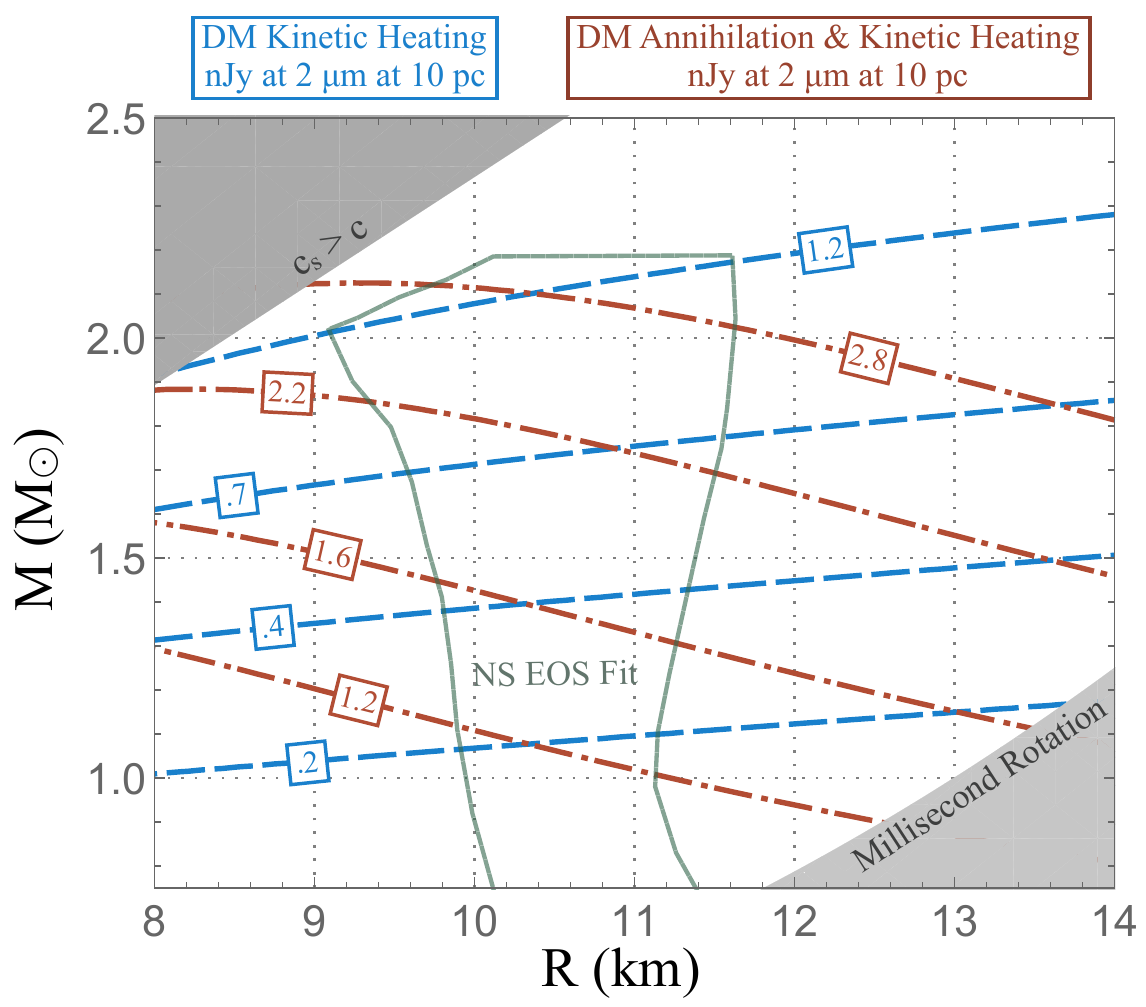}
\caption{Contours of infrared photon spectral flux density (in nanoJansky) are given for a neutron star ten parsecs from Earth with mass $M$ and radius $R$, maximally heated by dark matter with local density $0.42 ~{\rm GeV ~cm^{-3}}$ and relative speed $v_{\rm x} \sim 230~{\rm km ~s^{-1}}$. Dashed blue contours indicate dark kinetic heating only, while dotted-dashed red contours indicate heating from dark matter that also annihilates in the neutron star. The green region encloses a fit to pulsar data \cite{Ozel:2015fia}; gray regions are excluded by causality and the fastest rotating pulsars \cite{Lattimer:2006xb}.}
\label{fig:darklum}
\end{figure}

The minimum temperature of neutron stars induced by dark kinetic heating will depend on the ambient dark matter density and emission properties of the neutron star at low temperatures.  The heating of neutron stars by dark matter is similar to heating by the interstellar medium \cite{Zeldovich:1969aa}: in both cases, a particle crosses the neutron star surface, scatters against nucleons, and thereby transfers its kinetic energy to nucleons in the star. In cold neutron stars, the crust and core thermalize within less than a year after energy injection \cite{Ushomirsky:2001pd}. Following thermalization of scattered nucleons in the neutron star interior, dark kinetic heating would impart a minimum apparent neutron star luminosity (at long distances) of
\begin{align}
L_{\infty}^{\rm dark} = \dot{E}_{\rm k} \left(1-\frac{2GM}{R} \right) = 4 \pi \sigma_{\rm B} R^2 T_{\rm s}^4 \left(1-\frac{2GM}{R} \right),
\end{align}
where $\sigma_{\rm B}$ is the Stefan-Boltzmann constant, and $T_{\rm s}$ is the blackbody temperature of the neutron star as would be seen at its surface. For a near-Earth dark matter mass flux of $\dot{m} \sim 4 \times 10^{25}~{\rm GeV~s^{-1}}$, the neutron star will appear as a blackbody with an apparent temperature of up to $T_{\infty}^{\rm dark} \sim 1750~{\rm K}$, depending on the fraction of dark matter captured. For neutron stars inside the galactic bulge, maximal dark kinetic heating produces optical emission, $T_{\infty}^{\rm dark} \sim 3850~{\rm K}~ (\rho_{\rm x} / 10~{\rm GeV~cm^{-3}})^{1/4}$. In Figure \ref{fig:kindm}, we show the dark matter - neutron cross-section sensitivity obtained by finding $100-1750$ K neutron stars near the solar position where $\rho_{\rm x} \simeq 0.42 {\rm~GeV~cm^{-3}}$ \cite{Pato:2015dua}.

\emph{{\bf 4.} Detecting dark kinetic heating.} 
The locations of nearby neutron stars can be determined by detecting their radio pulses. While the radio observability of isolated, $\sim$Gyr old pulsars is a topic of active research, some of the faintest, oldest pulsars are likely to be uncovered by the recently operational FAST radio telescope \cite{2011IJMPD..20..989N}. After radio detection, infrared telescopes trained on old neutron stars can measure or bound their thermal emission. For a local dark matter density, the neutron stars will be heated to $T_{\infty}^{\rm dark} \sim 1750~{\rm K}$ by dark kinetic heating, with a spectrum peaking around $1-2 \ \mu$m. More precisely the blackbody spectral flux density of the neutron star is
\begin{align}
f_{\nu} &= \pi B(\nu, 
  T_\infty^{\rm dark})\times \frac{4\pi (R \gamma)^2}{4\pi d^2}~,
\end{align}
where
$
 B(\nu, 
  T) = 4 \pi \nu^3 \left( e^{\frac{2 \pi \nu}{k_b T}}-1 \right)^{-1}.
$
Figure \ref{fig:darklum} displays the dark kinetic heating spectral flux density for a range of neutron star masses and radii. At an observation distance $d = 10 \,\mathrm{pc}$ (where 1 - 5 old, cold neutron stars should abide \cite{1993ApJ...403..690B}) and at wavelength $\nu^{-1} = 2~ \mu$m, this results in a spectral flux density of $f_{\nu} \simeq 0.5~\mathrm{nanoJansky~(nJy)}$, which is near the optimal sensitivity of upcoming infrared telescopes, both space- and ground-based: the James Webb Space Telescope (JWST), Thirty Meter Telescope (TMT), and European Extremely Large Telescope (E-ELT). JWST, the telescope nearest to completion, is expected to reach $\mathcal{O}(10)$ signal-to-noise (SNR) for $\mathcal{O}(10 \mathrm{~nJy})$ in typical integration times of $10^4$ seconds. 

In more detail,the NIRCam on JWST is a 0.6 to 5 micron imager. There is a smorgasbord of filters available for NIRCam~\cite{JWST}; the F200W filter, centered at $2~\mu$m obtains $7.9$ nJy at 10 SNR in $10^4$ seconds, where for such long exposures, added sensitivity to spectral flux density scales with the square root of integration time. Using these values, a neutron star at distance $d$, maximally heated by dark matter kinetic energy (to 0.5 nJy at 2 $\mu$m), could be detected at SNR 2 in $ \sim 10^5\left({d}/{(10~\mathrm{pc})}\right)^4$ seconds. At the TMT, the IRIS instrument \cite{Wright:2010pk} has coverage from 0.8 to 2.5~$\mu$m.
 Although it has a larger collecting area than JWST, TMT must contend with the background provided by Earth's atmosphere. TMT's IRIS instrument has a projected K-band ($2.0$ to $2.4~\mu$m) sensitivity that permits SNR 2 detection of the aforementioned neutron star in $\sim 7 \times 10^4\left({d}/{(10~\mathrm{pc})}\right)^4$ seconds.

Much less integration time is required to detect neutron stars warmed by both dark kinetic heating and dark matter annihilating at their cores. Furthermore, since its sensitivity peaks at shorter wavelengths, the TMT does appreciably better at detecting neutron stars that emit higher energy photons, for which the infrared atmospheric background is reduced. A $10$~pc distant neutron star maximally warmed to a blackbody temperature of $T_{\infty}=$ 2480 K ($1.7$ nJy at $2~\mu$m) by dark matter annihilation and kinetic heating can be resolved in $9000 ~{\rm s}$ with the F200W filter in NIRCam at JWST versus $2000~{\rm s}$ in Y-band ($0.9$ to $1.2~\mu$m) with the TMT IRIS instrument.

\begin{figure}[t!]
\includegraphics[width=0.49\textwidth]{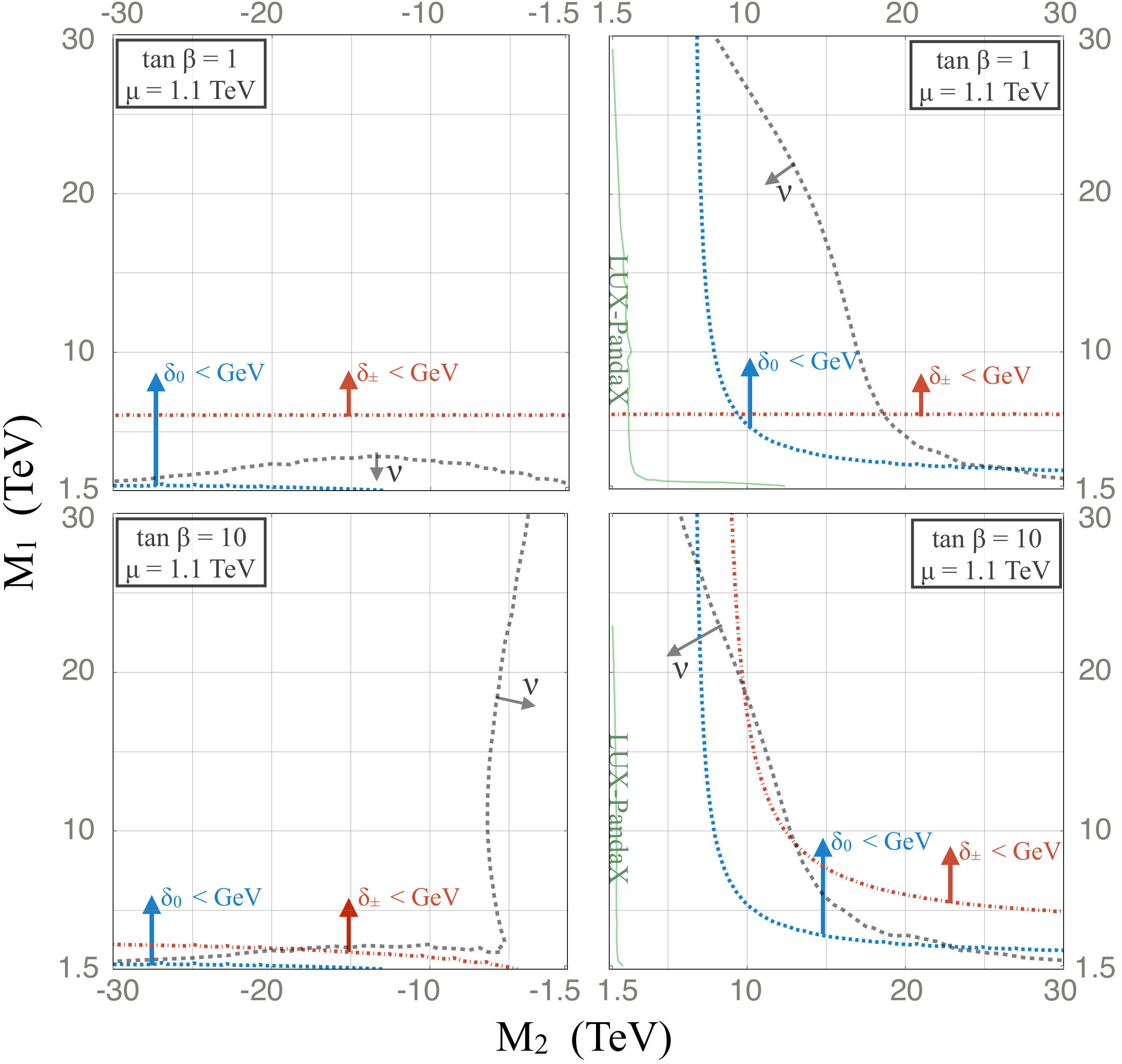}
\caption{Electroweakino mass parameters $M_1,M_2$ are shown with Higgsino mass $\mu = 1.1 ~{\rm TeV}$, which results in a freeze-out relic abundance of Higgsinos matching observations \cite{Hryczuk:2010zi}. Regions where either the neutral ($\delta_0$) or charged ($\delta_{\pm}$) inelastic mass splitting is less than one GeV would be uncovered by the first observation (or non-observation) of neutron star dark kinetic heating.  These regions are indicated above the blue dotted (red dot-dashed) lines where $\delta_0 < $ GeV ($\delta_\pm < $ GeV), permitting tree-level $Z$ boson ($W$ boson) exchange between semi-relativistic Higgsinos and neutrons in the neutron star. Regions below the black dotted line would be ruled out by direct detection searches sensitive to cross-sections as small as the xenon direct detection neutrino floor. LUX and PandaX presently bound pure Higgsinos below the green solid line for $M_2 > 0$, and have no sensitivity to parameters shown for $M_2 < 0$. Spin-independent cross-sections were computed using \toolfont{SuSpect} \cite{Djouadi:2006bz} and \toolfont{micrOMEGAs} \cite{Belanger:2014vza} with non-neutralino supersymmetric mass parameters set to $8$ TeV, as in \cite{Bramante:2015una}.}
\label{fig:darkinsinos}
\end{figure}

Finally, while so far we have focused on neutron stars near Earth, it is plausible to consider detecting neutron stars $\gtrsim 50$ parsecs from Earth, using longer integration times. At 50 parsecs, which is near the distance to a number of known pulsars, TMT using the IRIS Y-band would detect neutron star thermal emission arising from maximal dark matter annihilation and kinetic heating after $\sim 10^6$ s. Large next-generation telescopes have the benefit of adaptive optics, which result in an integration time that decreases as the fourth power of telescope diameter ($t_{\rm int} \propto D^{-4}$) for background-dominated surveys. This scaling holds when comparing IRIS's H-band sensitivity \cite{TMTHAndbook} to MICADO's \cite{2010SPIE.7735E..2AD}. Extrapolating to a Y-band sensitivity at E-ELT, this implies that an annihilation-heated, 70 parsec-distant neutron star can be detected in $\lesssim 10^6~ {\rm s}$. Likewise, E-ELT detects a 25 parsec-distant neutron star warmed only by dark kinetic heating in $\lesssim 10^6 $ s. Depending on the neutron star's position relative to Earth, it may be possible to combine a lengthy neutron star observation with a deep field survey; we leave a proposal along these lines to future investigation.

\emph{{\bf 5.} Parsing dark matter models with dark heating.} 
An old neutron star observed to have $T < 1750$ K would bound dark matter-nucleon cross-sections (see Figure~\ref{fig:kindm}). On the other hand, if a $T \lesssim 2000$ K population of neutron stars begins to emerge, dark kinetic heating can be used both to characterize the dark sector and to begin confirming that neutron stars are heated by dark matter as opposed to a Standard Model heating process. 

The energy injected by dark kinetic heating alone has a different dependence on the neutron star's mass and radius than the energy injected by dark matter that also annihilates at the core of the neutron star. This follows from comparing the maximum dark kinetic heating rate, $ (\gamma -1) \dot{m} $, with the maximum annihilation and dark kinetic heating rate, $ \gamma \dot{m} $. Note that $\dot{m}$ and $\gamma$ depend on the neutron star radius and mass, and these dimensions range over $R \sim 10-12$ km \cite{Steiner:2012xt,Ozel:2015fia} and $M \sim 1-2 M_{\odot}$~\cite{Kiziltan:2013oja}. Consequently, annihilating dark matter will heat populations of old neutron stars to luminosities that range over a factor of $\lesssim 3$, while dark matter that only deposits kinetic energy will heat neutron stars to luminosities ranging over a factor of $ \sim 10-15$. This can be seen in Figure~\ref{fig:darklum}, which shows the spread in luminosity from dark kinetic heating as a function of neutron star mass and radius, both with and without dark matter annihilation in the star. This methodology can also be used in future observations, to differentiate dark matter annihilation in neutron stars, from neutron stars warmed by interstellar medium accretion -- the spread in neutron star luminosities will be broader for the latter scenario.

\emph{{\bf 6.} An infrared window on pure Higgsinos.}
We now demonstrate, using an explicit model of Higgsino dark matter, that neutron star dark kinetic heating could reveal inelastic dark matter otherwise inaccessible to direct detection experiments. Neutralinos are the spin $\frac{1}{2}$ superpartners of electroweak bosons, ``electroweakinos," and Higgs bosons, ``Higgsinos" \cite{Martin:1997ns,Feng:1999fu,Gherghetta:1999sw}. After electroweak symmetry breaking, the neutral components of the Higgsinos and electroweakinos mix as a result of Higgs-Higgsino-electroweakino couplings. If the electroweakino mass parameters $M_1$ (``bino") and $M_2$ (``wino") are much larger than the Higgsino mass parameter $\mu$, the two lightest neutralino mass eigenstates ($\chi^0_1,\chi^0_2$) and the lightest charged mass eigenstate ($\chi^\pm_1$) will be mostly Higgsino.

Pure Higgsinos ($m_{\rm Z} \ll |\mu| \ll |M_1|,|M_2| \lesssim 10^7~{\rm GeV} $) scatter elastically with nucleons  ($\chi^0_1 n \rightarrow \chi^0_1 n$) at terrestrial direct detection experiments through weak boson loops with a small cross-section, $\sigma_{\rm n x} \lesssim 10^{-48} ~{\rm cm}^2$ \cite{Hisano:2011cs,Hill:2013hoa}, which is beyond the reach of any planned direct detection experiment \cite{ArkaniHamed:2006mb,Cheung:2012qy,Bramante:2015una}. This can be compared to pure Higgsino inelastic processes $\chi^0_1 n \rightarrow \chi^0_2 n$ and $\chi^0_1 n \rightarrow \chi^\pm_1 p^\mp$, which proceed through tree-level exchange of $Z$ and $W$ bosons with a much larger nucleon scattering cross-section, $\sigma_{\rm n \rm x} \sim 10^{-39}~{\rm cm^2}$ \cite{Bramante:2016rdh}. These tree-level weak boson exchange scattering processes are forbidden if the recoil energy exchange is less than the inelastic mass splitting, $e.g.$ the $Z$-exchange is forbidden if $\delta_0 \equiv m_{\chi^0_2}- m_{\chi^0_1} > \Delta E_{\rm s}$ \cite{Bramante:2016rdh,Nagata:2014wma}. Here we will show that nearly all pure Higgsino parameter space inaccessible to direct detection experiments is accessible with dark kinetic heating, because inelastic inter-state transitions up to $\delta \sim$ GeV are permitted for semi-relativistic dark matter scattering against neutron stars, compared with $\delta \lesssim 500$ keV at direct detection experiments \cite{Bramante:2016rdh}.

The neutral and charged Higgsino components ($\chi^0_1,\chi^0_2, \chi_1^\pm$) are split in mass by mixing with electroweakinos and by electroweak loop corrections \cite{Fritzsche:2002bi,Oller:2003ge},
\begin{align}
 \delta_0 &\simeq \frac{v^2}{4}\left(\frac{g^2_1}{M_1} + \frac{g^2_2}{M_2}\right), \nonumber \\
\nonumber \delta_\pm^{\rm tree} &\simeq \frac{v^2}{4} \left(\frac{g_1^2}{M_1} (1+\sin2\beta) + \frac{g_2^2}{M_2}(1-\sin2\beta) \right),\\
\delta_\pm^{\rm loop} &\simeq \left(\frac{g_2}{4\pi}\right)^2 \mu \sin^2\theta_W f\left(\frac{m_Z}{\mu}\right),
\label{eq:charsplit}
\end{align}
where $\tan \beta$ is the ratio of the Higgs vevs and 
$
\nonumber f(x) = 2 \int^1_0 dt (1+t) \log\left(1 + \frac{x^2(1-t)}{t^2}\right) 
\xrightarrow{x \to 0} 2\pi x - 3 x^2~.
$
Thus, for $\mu \gg m_Z$, $\delta_\pm^{\rm loop} \simeq$ 355 MeV.

Dark kinetic heating occurs for regions where either $\delta_0$ or $\delta_\pm = \delta_\pm^{\rm tree} + \delta_\pm^{\rm loop}$ are $< 1~$GeV, which allows for tree level scattering via $Z$ or $W$ exchange. Fig.~\ref{fig:darkinsinos} shows that for most parameter space outside the reach of future direct detection experiments, Higgsino mass splittings are sub-GeV, implying that all Higgsinos passing through the neutron star would deposit their kinetic energy. Whether Higgsinos thermalize \cite{Bertoni:2013bsa} and annihilate in the neutron star core depends on the low-energy Higgsino-nucleon cross-section, which requires further study \cite{Hill:2013hoa}.

\emph{{\bf 7.} Concluding with strong interactions.}
We have pointed out that dark matter's interactions with Standard Model particles could be unmasked by looking for infrared thermal emission from neutron stars near the solar position. This effect applies to many dark matter models, including weakly interacting (WIMP), pure Higgsino, and strongly-interacting (SIMP) dark matter. For SIMPs \cite{Rich:1987st,Dimopoulos:1989hk,Starkman:1990nj,Albuquerque:2003ei,Erickcek:2007jv,Mack:2007xj} with up to nuclear cross-sections, the sensitivity shown in Figure \ref{fig:kindm} extends to $m_{\rm x} \sim 10^{27}~{\rm GeV}$. For super-nuclear cross-sections, some dark matter scattering occurs near enough to the neutron star surface that high energy photons are produced alongside dark kinetic infrared blackbody emission \cite{Zeldovich:1969aa}. SIMP-scattered high energy photons, dark kinetic X-rays from neutron stars near Sagittarius A*, and other exciting applications of dark kinetic heating will be explored in future work.

\acknowledgments We are especially indebted to Suresh Sivanandam for extensive guidance on infrared astronomy. It is a pleasure to thank Asimina Arvanitaki for comments on the manuscript, and Katie Auchettl, Sergei Dubovsky, Fatemeh Elahi, Robert Lasenby, Adam Martin, and Maxim Pospelov for useful conversations. We thank exoplanet scientists in advance for hastening the construction of a 100 meter telescope. Research at Perimeter Institute is supported by the Government of Canada through Industry Canada and by the Province of Ontario through the Ministry of Economic Development \& Innovation. T.~L. is supported by NSF Grant PHY-1404311 to John Beacom. N.~R. is supported by NSF Grant PHY-1417118. 

\bibliography{darkinetic}

\end{document}